\documentclass[aps,prb,twocolumn]{revtex4}

\usepackage{graphicx}
\usepackage{dcolumn}
\usepackage{bm}
\usepackage{float}
\usepackage{amsfonts}
\usepackage{epsfig}

\begin{document}

\preprint{APS/XYZ}

\title{Stability and structure of atomic chains on Si(111)}

\author{Corsin Battaglia}
\affiliation{Institut de Physique, Universit\'e de Neuch\^atel,
2000 Neuch\^atel, Switzerland}%
\email{corsin.battaglia@unine.ch}
\homepage{http://www.unine.ch/phys/spectro}
\author{Steven C. Erwin}
\affiliation{Center for Computational Materials Science, Naval
Research Laboratory, Washington D.C. 20375, USA}
\author{Philipp Aebi}
\affiliation{Institut de Physique, Universit\'e de Neuch\^atel,
2000 Neuch\^atel, Switzerland}%

\date{\today}

\begin{abstract}
  We study the stability and structure of self-assembled atomic chains on
  Si(111) induced by monovalent, divalent and trivalent adsorbates,
  using first-principles total-energy calculations and
  scanning tunneling microscopy. We find that only structures
  containing exclusively silicon honeycomb or silicon Seiwatz
  chains are thermodynamically stable, while mixed configurations,
  with both honeycomb and Seiwatz chains, may be kinetically stable.
  The stability and structure of these atomic chains can be understood
  using a surprisingly simple electron-counting rule.
\end{abstract}

\pacs{} \keywords{}

\maketitle

\section{Introduction}
\begin{figure}
\centerline{\psfig{file=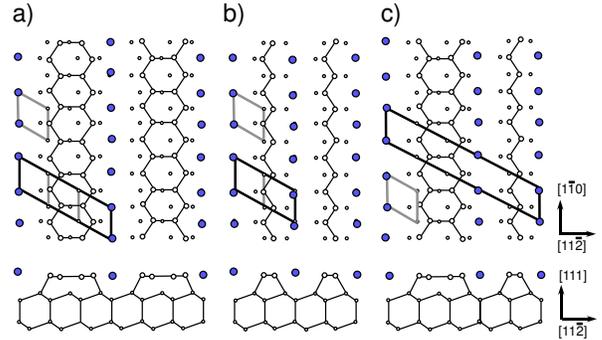,width=0.9\linewidth,angle=0}}
\caption{\label{fig:Model}(Color online) Top (upper) and side
(bottom) view of the two prototypical silicon chains separated
by channels that accommodate the adsorbate atoms (filled blue
circles). (a) Honeycomb chain with (3$\times$1) unit cell. (b)
Seiwatz chain with (2$\times$1) unit cell. (c) Mixed chain
structure with (5$\times$1) unit cell. The (1$\times$1)
unit cell of the (111)-oriented substrate is indicated in grey. The
crystallographic directions of the substrate are also indicated.}
\end{figure}

Understanding and controlling the structure of surfaces on the
atomic level is of tremendous technological importance. This is
especially true for the growth of semiconductor nanostructures,
where the competition between thermodynamics and kinetics can play a
decisive role. The reconstruction of semiconductor surfaces is
driven by the elimination of dangling bonds and the minimization of
surface stress, with a striking diversity of outcomes. Despite this
variety, even very elaborate architectures are generally comprised
of a small number of elementary structural building blocks. Dimers
and adatoms on Si(100) and Si(111), respectively, are the best known
strategies for reducing the number of dangling bonds. Tetramers and
pentamers, encountered on Si(114) \cite{Erwin96}, Si(113)
\cite{Laracuente98}, Si(110) \cite{An00,Stekolnikov04b}, and Si(331)
\cite{Battaglia08} constitute more complex units. For
adsorbate-induced surface reconstructions of the Si(111) surface,
honeycomb \cite{Collazo-Davila98,Lottermoser98,Erwin98} and Seiwatz
chains \cite{Seiwatz64} have recently emerged as universal building
blocks \cite{Battaglia07} (see Fig. \ref{fig:Model}). These form the
basis of a large class of atomic chain reconstructions which have
been the focus of intense research because of their fascinating
quasi one-dimensional electronic properties
\cite{Crain03,Ahn05,Guo05,Snijders06}. The fact that only silicon
atoms participate in the formation of the honeycomb and Seiwatz
chains means that a number of different adsorbates can induce a
chain reconstruction, simply by donating the correct number of
electrons to the substrate.

We have recently revealed remarkable systematics in this class of
adsorbate-induced reconstructions, relating the valence state of the
adsorbate to the allowed coverages and periodicities of the
resulting adsorbate chains \cite{Battaglia08b}.  All experimentally
observed phases satisfy a simple electron-counting rule: the
adsorbates must provide either one electron (to stabilize a
(3$\times$1) honeycomb-chain unit), or two electrons (to stabilize a
(2$\times$1) Seiwatz-chain unit).\cite{Battaglia07} For monovalent
adsorbates the only experimentally observed configuration is
obtained with an adsorbate coverage of 1/3 ML.\cite{footnote1} This
corresponds to one adsorbate donating one electron per (3$\times$1)
honeycomb chain unit as shown in Fig. \ref{fig:Model}(a) , in
agreement with the electron-counting rule. For divalent adsorbates,
only 1/6 ML (half the monovalent adsorbate coverage) is required to
stabilize honeycomb chains, since a divalent adsorbate donates two
electrons. The pure Seiwatz chain structure shown in Fig.
\ref{fig:Model}(b) is stable for 1/2 ML of divalent adsorbates,
resulting from two donated electrons per (2$\times$1) Seiwatz chain
unit. A mixed chain phase, alternating between honeycomb and Seiwatz
chains as shown in Fig. \ref{fig:Model}(c), observed for divalent
adsorbates, at intermediate coverage of 3/10 ML, also satisfies the
electron-counting rule. Here the divalent adsorbates supply one
electron to the honeycomb chain unit and two electrons to the
Seiwatz chain unit.  Similarly, trivalent adsorbates at a coverage
of 2/10 ML donate the same number of electrons, allowing the same
mixed phase to be stabilized.

In this report we subject these observed systematics, and the
electron-counting rule deduced from them, to more detailed
theoretical scrutiny. Specifically, we examine theoretically the
thermodynamic and kinetic stability of several chain reconstructions
of Si(111) using first-principles total-energy methods. We compare
the resulting pictures that emerge for three prototypical
adsorbates: monovalent (Na), divalent (Ca), and trivalent (Gd).  For
each adsorbate we compare a number of reconstructions with different
adsorbate coverage, including the bare Si(111)-(7$\times$7)
reconstruction and, for Ca and Gd adsorbates, the silicide phase
experimentally observed at higher coverage.  For each candidate
reconstruction we determine the surface energy as a function of
adsorbate chemical potential, and thereby determine the energy
ordering of our candidate reconstructions at any thermodynamically
allowed value of chemical potential. We find that the
thermodynamically stable chain reconstructions are formed
exclusively from either honeycomb or Seiwatz chains, with mixed
phases slightly higher in energy. We also argue that for Gd
adsorbates the experimentally observed mixed phase, which combines
honeycomb and Seiwatz chains, while not thermodynamically stable, is
kinetically stable.

\section{Methods}
We used first-principles total-energy calculations to determine
equilibrium geometries and relative surface energies.  The
calculations were performed in a slab geometry with up to six layers
of Si plus the reconstructed surface layer. All atomic positions were
relaxed except for the bottom layer, which was passivated. Total
energies and forces were calculated within the generalized-gradient
approximation to density-functional theory (DFT), using
projector-augmented-wave (PAW) potentials
\cite{kresse_phys_rev_b_1993a,kresse_phys_rev_b_1996a}. We checked that
the slab thickness, plane-wave cutoff, and sampling of the surface
Brillouin zone were each sufficient to converge the relative surface
energies to within 1 meV/\AA$^2$.

For calculations with Gd adsorbates, the seven 4$f$ electrons were
treated explicitly as valence states. The possibility of magnetic
order among Gd atoms within a single fully occupied channel was
investigated in one case, with the result that ferromagnetic
ordering was slightly preferred, by 0.1 eV, to antiferromagnetic
ordering. Based on this finding we assumed ferromagnetic order for
all Gd phases. We also found that putting the 4$f$ electrons in the
core led to only insignificant changes to the calculated absolute
surface energies. This establishes that magnetic order among the Gd
atoms plays no substantive role in the stability of different
surface phases.

Growth and STM experiments were carried out in a ultra-high vacuum
chamber with a residual gas pressure of $3\times10^{-11}$ mbar
equipped with an Omicron LT-STM. Boron-doped Si(111) with a
resistivity of 5 $\Omega\cdot$cm was heated by passing a direct
current. Gd was evaporated from a water-cooled $e$-beam evaporator.

We consider first the stability of atomic chains induced by
monovalent adsorbates such as the alkali metals Li, Na, K, Rb, and
Cs.  These are known to induce a chain reconstruction on Si(111)
exhibiting simple (3$\times$1) periodicity. The widely accepted
structural model, the so-called honeycomb chain-channel (HCC) model
\cite{Collazo-Davila98,Lottermoser98,Erwin98}, is shown in Fig.
\ref{fig:Model}(a).  The HCC model consists of Si honeycomb chains
aligned along the $[1\overline{1}0]$ direction, separated by empty
channels. The adsorbate atoms occupy sites within these channels.

The presence of an adsorbate, such as Na, in the HCC model does not
allow direct comparison between its surface energy and the surface
energy of clean reconstructed Si(111)-(7$\times$7). The proper way
to compare energies of structures differing in stoichiometry is via
the chemical potentials $\mu_{\rm Si}$ and $\mu_{\rm Na}$ of the
constituents \cite{kaxiras1987a}, which are the energy per atom
available in the reservoirs with which the surface is assumed to be
in equilibrium. The surface energy (per unit area) is then
\begin{equation}
\gamma  =  E_{\rm surf}/A  = (E_{\rm tot}
- n_{\rm Si}\mu_{\rm Si} - n_{\rm Na}\mu_{\rm Na})/A,
\label{gamma1}
\end{equation}
where $E_{\rm tot}$ is the total energy of a double-sided slab whose
unit cell, with total area $A$, contains $n_{\rm Si}$ Si atoms and $n_{\rm Na}$ Na atoms.
Since the surface is in equilibrium with the bulk Si substrate,
$\mu_{\rm Si}$ is the energy per atom in bulk Si. The adsorbate
chemical potential, $\mu_{\rm Na}$, however, corresponds to a real
physical variable that can be externally tuned by, for example,
varying the partial pressure of Na.  Intuitively, increasing the Na
partial pressure will increase the stability of structures with higher
Na coverage.
This is evident from recasting Eq.~\ref{gamma1} as
\begin{equation}
\gamma  =  \gamma_0 - \theta_{\rm Na}\mu_{\rm Na},
\label{gamma2}
\end{equation}
where $\theta_{\rm Na}$ is the adsorbate coverage. From
Eq.~\ref{gamma2} it is clear that reconstructions with larger
$\theta_{\rm Na}$ are increasingly favored as $\mu_{\rm Na}$
increases.  For a given value of $\mu_{\rm Na}$ the reconstruction
with the lowest surface energy $\gamma$ will be realized in an
experiment, if it is performed under conditions of thermodynamic
equilibrium. Phase transitions can thus occur as $\mu_{\rm Na}$ is
changed.

Thermodynamics places an upper bound on the adsorbate chemical
potential, $\mu_{\rm Na}\leq\mu^{\rm 0}_{\rm Na}$, given by the energy
per atom in the ground-state (body-centered cubic) phase of
elemental Na.  Exceeding this limit in an experiment would result in
precipitation of elemental Na, because that phase would then be
energetically preferable to any adsorbed phase.  When making the
chemical potential sufficiently low, by turning the partial pressure
to a very small value, the bare surface will be the most stable phase.
The more interesting question is what happens in between these two
extremes.

\begin{figure}
\includegraphics{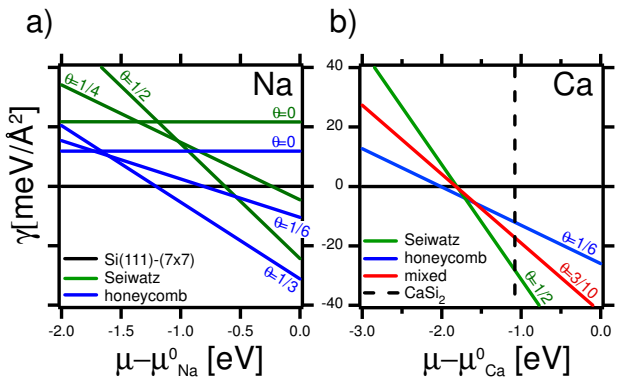}%
\caption{\label{fig:Na}(Color online) (a) Surface-energy diagram for
monovalent Na adsorbates. We compare pure honeycomb chains
($\theta$=1/3, 1/6 and 0 ML, blue lines) and pure Seiwatz chains
($\theta$=1/2, 1/4 and 0 ML, green lines). The surface energy of
Si(111)-(7$\times$7) ($\theta$=0 ML, black line) is also shown. (b)
Surface-energy diagram for divalent Ca adsorbates. We compare pure
honeycomb chains ($\theta$=1/6, blue line), pure Seiwatz chains
($\theta$=1/2, green line) and a mixed chain configuration with
alternating honeycomb and Seiwatz chains ($\theta$=3/10, red line).
The vertical dashed line represents the bulk CaSi$_2$ silicide. }
\end{figure}

\section{Results \& Discussion}
\subsection{Monovalent adsorbates}
The calculated DFT surface energies, as a functional of chemical
potential, are shown in Fig. \ref{fig:Na}(a) for several
reconstructions of Si(111) induced by the monovalent adsorbate Na.
All surface energies are given relative to that of the bare
Si(111)-(7$\times$7), which we place at $\gamma$=0.\cite{footnote2}
 The colored lines with non-zero slopes represent the surface
energies for various Na-induced reconstructions.  Results for two
HCC phases are shown (blue lines), with $\theta_{\rm Na}$=1/3 and
1/6 ML corresponding to fully occupied (3$\times$1) and half-filled
(3$\times$2) channels, respectively. Also shown are results for two
Seiwatz-chain phases (green lines), with $\theta_{\rm Na}$=1/2 and
1/4 ML, corresponding to fully occupied (2$\times$1) and half-filled
(2$\times$2) channels, respectively. Finally, results for ``empty''
HCC and Seiwatz reconstructions (i.e.~without adsorbates) are shown
as flat lines.

From the energy ordering of these various reconstructions, it is
evident that over the allowed range of $\mu_{\rm Na}$ only two
phases are thermodynamically stable: the clean Si(111)-(7$\times$7)
surface and the (3$\times$1) HCC phase with $\theta_{\rm Na}$=1/3,
in agreement with experiment. We find, but do not here show, very
similar results for other monovalent adsorbates (Li, K), and
conclude that for monovalent adsorbates the only thermodynamically
stable phase is the HCC with every site in the channel occupied.

\subsection{Divalent adsorbates}
We turn now to reconstructions induced by divalent adsorbates, such
as the alkaline earth metals (Mg, Ca, Sr, Ba) and rare earth metals
(Sm, Eu, Yb). From experiments we know that at a coverage of 1/6 ML,
divalent adsorbates stabilize (3$\times$"2") honeycomb chains. At a
coverage of 1/2 ML, the (2$\times$1) Seiwatz chains are
experimentally observed.\cite{Baski01,Sekiguchi01} At intermediate
coverages, divalent adsorbates are known to stabilize mixed chain
structures with higher periodicity consisting of a combination of
honeycomb chains and Seiwatz chains. The simplest combination
alternates between honeycomb and Seiwatz chains, resulting in
(5$\times$1) unit cell, as shown in Fig.\ \ref{fig:Model}(c). The
(5$\times$"2") periodicity observed for divalent adsorbates may be
thought of as being built from two (3$\times$1) honeycomb-chain
units and two (2$\times1$) Seiwatz-chain units.

We have calculated the surface energies of a variety of candidate
reconstructions based on the divalent adsorbate Ca, including pure
honeycomb chains with coverages $\theta$=1/3 and 1/6 ML, pure Seiwatz
chains with coverages $\theta$=1/2 and 1/4 ML, and the mixed
configuration alternating between honeycomb and Seiwatz chains with
adsorbate coverages $\theta$=2/10, 3/10, and 4/10 ML. A new
consideration arises for Ca, because Si and Ca can form a variety of
stable bulk silicides, such as CaSi$_2$ and Ca$_2$Si. To prevent
precipitation of these bulk phases, the Ca and Si chemical potentials must
also satisfy the inequalities
\begin{eqnarray}
\mu_{\rm Ca}  +  2\mu_{\rm Si} &  \leq  & \mu({\rm CaSi_2}),\\
2\mu_{\rm Ca} +   \mu_{\rm Si} &  \leq  & \mu({\rm Ca_2Si}),
\end{eqnarray}
where $\mu({\rm CaSi_2})$ is the energy per formula unit of CaSi$_2$,
and likewise for Ca$_2$Si. Since $\mu_{\rm Si}$ is fixed, these
constraints have the effect of further lowering the highest allowed
value of $\mu_{\rm Ca}$.

The resulting surface energies for Ca-induced reconstructions are
shown in Fig.\ \ref{fig:Na}(b). To keep the figure uncluttered we
have only plotted the phases that are stable, or close to stable,
for some allowed value of $\mu_{\rm Ca}$.  Two Ca-induced phases are
thermodynamically stable: the HCC reconstruction with $\theta$=1/6
ML (blue) and the Seiwatz-chain reconstruction with $\theta$=1/2 ML
(green).  The mixed HCC+Seiwatz configuration with $\theta$=3/10 ML
(red line) is energetically just above these two phases, but passes
so close (within 1 meV/\AA$^2$) to their intersection point that its
formation cannot be ruled out.  Experimentally, the mixed
configuration is indeed found at coverages between those of the two
pure phases.

 \begin{figure}
\includegraphics{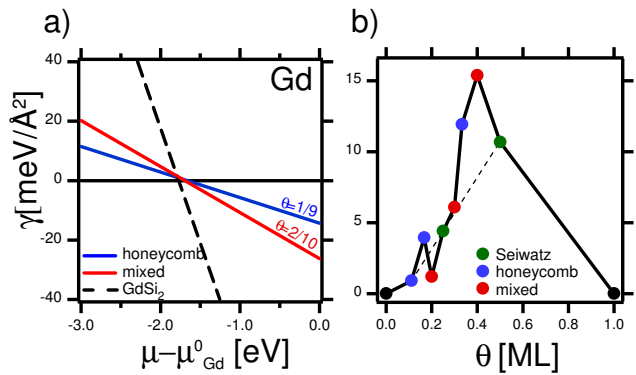}%
\caption{\label{fig:Gd}(Color online) (a) Surface-energy diagram for the trivalent
  adsorbate Gd. We compare pure honeycomb chains ($\theta$=1/9, blue)
  and a mixed chain configuration with alternating honeycomb and
  Seiwatz chains ($\theta$=2/10, red).  The horizontal line represents
  Si(111)-(7$\times$7) ($\theta$=0 ML, black), and the dashed
  line represents the GdSi$_2$ silicide film ($\theta$=1 ML). (b)
  Surface energies of Gd phases plotted as a function of coverage.
  Only the points have meaning; the lines are added for clarity.}
\end{figure}

\subsection{Trivalent adsorbates}
Experiments using trivalent adsorbates (Gd, Dy, Er, Ho) reveal a
mixed configuration with (5$\times$"2") periodicity consisting of
alternating honeycomb and Seiwatz chains, as shown in Fig.
\ref{fig:Model}(c). Pure honeycomb or pure Seiwatz chain structures
are not observed. We have calculated the surface energies of a
number of hypothetical Gd-induced configurations, including pure
honeycomb chains with every channel site occupied ($\theta$=1/3 ML),
every second site occupied ($\theta$=1/6 ML), and every third site
occupied ($\theta$=1/9 ML), as well as of Seiwatz chains with every
channel site occupied ($\theta$=1/2 ML), and every second site
occupied ($\theta$=1/4 ML). We also considered three mixed
configurations with $\theta$=2/10, 3/10, and 4/10 ML.  Finally, we
also calculated the surface energy for the well-studied epitaxial
GdSi$_2$ silicide, which consists of 1 ML of Gd on Si(111) in the
so-called ``B-T4'' structure.\cite{Rogero04,Bonet05}

The resulting surface energies for Gd-induced reconstructions are
shown in Fig.\ \ref{fig:Gd}(a). As before, we plot only the phases
that are stable, or nearly so.  There are two thermodynamically
stable phases, the bare Si(111)-(7$\times$7), and the GdSi$_2$
silicide phase with $\theta_{\rm Gd}$=1. None of the Gd-chain
reconstructions is thermodynamically stable within DFT.

To investigate the possibility that the experimentally observed
phases are kinetically stable, we examine in more detail the
energetics of all phases with intermediate Gd coverage. In
particular, we are interested in the tendency of an adsorbate phase
with intermediate coverage $\theta_{Gd}$ to separate into two stable
phases, one with lower coverage $\theta_{Gd}^-$ and one with higher
coverage $\theta_{Gd}^+$.  Let $x$ denote the fraction of the total
surface area occupied by the higher coverage phase. Then the average
coverage of the two phases is simply $x\theta_{Gd}^++
(1-x)\theta_{Gd}^-$. If $x$ is chosen such that this average
coverage is equal to the coverage $\theta_{Gd}$ of the homogeneous
phase, then the surfaces energies of the homogeneous and
inhomogeneous phases can be compared directly, without the need for
specifying a chemical potential.  To make this comparison simple we
return to Eq. 2 and now regard the surface energy gamma as a
function of coverage $\theta_{Gd}$ for a fixed, arbitrary value of
chemical potential $\mu_{Gd}$. Although the surface energies for the
individual phases depend explicitly on $\mu_{Gd}$, the energy
difference between the two alternative scenarios (the homogeneous
phase versus the phase-separated phase) with the same average
coverage does not. For display purposes we choose a value for which
the two endpoint phases have equal surface energies.

The resulting surface energies, for all the Gd chain phases we have
considered, are plotted in Fig.\ \ref{fig:Gd}(b) relative to the
endpoint phases, as a function of Gd coverage. For intermediate Gd
coverages, the surface energy is always higher than at the
endpoints. Thus, a surface prepared with intermediate coverage will,
under conditions of thermodynamic equilibrium, phase separate into
an appropriate mixture of the two endpoint phases.  But what about
conditions under which thermodynamic equilibrium cannot be achieved?

There are good reasons to consider this scenario. The conversion of an
intermediate phase to a combination of bare (7$\times$7) and the
GdSi$_2$ silicide phase requires a considerable rearrangement of the
top several atomic layers; this restructuring may be kinetically
hindered. On the other hand, the structural differences among the
various Gd-chain phases are relatively minor: for example, all are
quasi-one dimensional with similar underlying building blocks. Thus,
the thermodynamically favored endpoint phases may be kinetically
inaccessible, even though equilibrium is achieved among the subset of
structurally similar chain phases with intermediate coverage.

\begin{figure}
\includegraphics{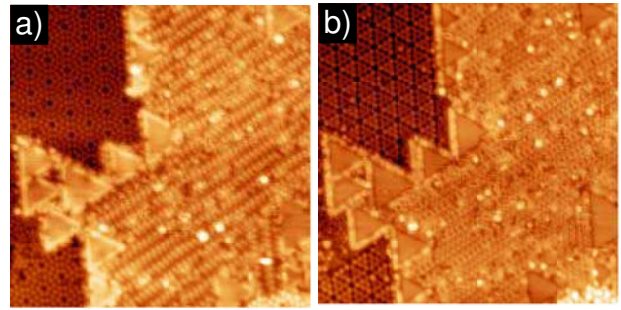}%
\caption{\label{fig:STM}(Color online) (a) Empty and (b)
filled-state STM image of Si(111) surface after Gd deposition and
subsequent annealing. Atomic chains coexist with clean areas with
the Si(111)-(7$\times$7) reconstruction separated by triangular
GdSi$_2$ islands. $U$=$\pm$1.8 V, $I$=0.18 nA, 45 nm$\times$45 nm. }
\end{figure}

To analyze the consequences of this hypothesis, we apply the Maxwell
construction to the phases with intermediate Gd coverage. This
geometrical analysis allows one to answer the question of which, if
any, intermediate phases are stable against phase separation into
appropriate mixtures of other phases. We consider again the results in
Fig.\ \ref{fig:Gd}(b) and construct an analogous plot (not shown),
this time with the (7$\times$7) and GdSi$_2$ endpoints excluded. Of
the remaining phases, only one is stable with respect to phase
separation into all possible other pairs: the (5$\times$2) mixed
HCC+Seiwatz phase with $\theta_{\rm Gd}$=2/10. This conclusion is easy
to visualize in Fig.\ \ref{fig:Gd}(b) by simply drawing straight lines
between all possible pairs of phases; the dotted line shows one such
example, demonstrating the stability of the $\theta_{\rm Gd}$=2/10
phase with respect to separation into the two endpoint phases, with
$\theta_{\rm Gd}$=1/9 and $\theta_{\rm Gd}$=1/2.

These findings are indeed consistent with our experimental
observations.  Starting with the deposition of 2/10 ML Gd onto the
substrate held at 680$^o$ C, followed by short annealing still at
680$^o$ C and a cool-down to room temperature, one obtains a surface
uniformly covered with atomic chains. After annealing for a slightly
longer period or at slightly higher temperatures, phase separation may
directly be observed in STM images (see Fig.  \ref{fig:STM}):
triangles of GdSi$_2$ silicide form together with regions of clean
Si(111)-(7$\times$7) coexisting with the chains.  Further annealing
completely transforms the chains
into silicide islands and areas with Si(111)-(7$\times$7).

\section{Conclusion}
We used first-principles total-energy calculations and scanning
tunneling microscopy to study the stability and structure of atomic
chains of monovalent, divalent, and trivalent adsorbates on Si(111).

For monovalent adsorbates the theoretical and experimental results
are in excellent agreement, both identifying the honeycomb-chain
channel (HCC) reconstruction with adsorbate coverage $\theta=$1/3 as
the only stable chain phase.

For divalent adsorbates three chain phases are found experimentally,
corresponding to coverages 1/6, 1/2, and intermediate values. These
findings are corroborated by our total-energy calculations, which
identify the three lowest-energy phases as a half-occupied HCC
phase, a fully occupied Seiwatz-chain phase, and a simple
combination of these.

For trivalent adsorbates, our total-energy calculations indicate
that the observed combination of HCC and Seiwatz phases with
adsorbate coverage 2/10 is kinetically stable with respect to phase
separation into other, thermodynamically stabler phases.

For all adsorbates, the thermodynamically (or kinetically) stable
phases all obey a surprisingly simple electron-counting rule
proposed earlier \cite{Battaglia07,Battaglia08b}.

\section*{Acknowledgments}
We thank Franz J. Himpsel for stimulating discussions. The help of
Leslie-Anne Fendt, Samuel Hoffmann, Claude Monney and Christoph
Walther is gratefully acknowledged. Skillful technical assistance
was provided by our workshop and electric engineering team. This
work was supported by the Fonds National Suisse pour la Recherche
Scientifique through Div. II, MaNEP and the US Office of Naval
Research. Computations were performed at the DoD Major Shared
Resource Center at ASC.

\bibliography{BiblioGdSi111Erwin}

\begin{thebibliography}{26}
\bibitem[{foo({\natexlab{a}})}]{footnote1}
\bibinfo{journal}{We define 1 ML as the atom density of the bulk-terminated
  Si(111) surface of $7.8\times10^{14}$ cm$^{-2}$}.
\bibitem[{foo({\natexlab{b}})}]{footnote2}
\bibinfo{journal}{The surface energy for the Si(111)-(7$\times$7)
  reconstruction was determined by calculating the surface energy of the Pandey
  model \cite{Pandey81} for Si(111)-(2$\times$1) and subtracting the published
  DFT energy difference of 60 meV per (1$\times$1) unit cell \cite{Brommer91}}.
\end{thebibliography}

\end{document}